\documentclass[aps,pre,reprint,superscriptaddress,showkeys,floatfix]{revtex4-2}
\usepackage[utf8]{inputenc}
\usepackage{amsmath}
\usepackage{amssymb}
\usepackage{amsfonts}
\usepackage{graphicx}
\usepackage{siunitx}
\usepackage{color}
\usepackage{soul}
\usepackage[colorlinks=true,citecolor=blue,urlcolor=red]{hyperref}

\newcommand{\fluc}{\mathrm{fluc}}
\newcommand{\cl}{\mathrm{cl}}
\newcommand{\mn}{\mathrm{min}}
\newcommand{\mx}{\mathrm{max}}
\newcommand{\diss}{\mathrm{diss}}
\newcommand{\avg}{\mathrm{avg}}

\newcommand{\meas}{\mathrm{meas}}
\newcommand{\theo}{\mathrm{theo}}
\newcommand{\NESS}{\mathrm{NESS}}
\newcommand{\EQ}{\mathrm{EQ}}
\newcommand{\Tfluc}{T^\fluc}

\newcommand{\B}{\mathrm{B}}
\newcommand{\D}{\mathrm{D}}
\newcommand{\eff}{\mathrm{eff}}

\renewcommand{\d}{\mathrm{d}}

\begin{document}

\title{Linking fluctuation and dissipation in spatially extended out-of-equilibrium systems}

\author{Alex Fontana}\affiliation{Univ Lyon, ENS de Lyon, Univ Claude Bernard Lyon 1, CNRS, Laboratoire de Physique, F-69342 Lyon, France}
\author{Ludovic Bellon}\email{Corresponding author: ludovic.bellon@ens-lyon.fr}\affiliation{Univ Lyon, ENS de Lyon, CNRS, Laboratoire de Physique, F-69342 Lyon, France}
\date{\today}

\begin{abstract}
For systems in equilibrium at a temperature $T$, thermal noise and energy damping are related to $T$ through the fluctuation-dissipation theorem (FDT). We study here an extension of the FDT to an out of equilibrium steady state: a microcantilever subject to a constant heat flux. The resulting thermal profile in this spatially extended system interplays with the local energy dissipation field to prescribe the amplitude of mechanical fluctuations. Using three samples with different damping profiles (localized or distributed), we probe this approach and experimentally demonstrate the link between fluctuations and dissipation. The thermal noise can therefore be predicted \emph{a priori} from the measurement of the dissipation as a function of the maximum temperature of the micro-oscillator.
\end{abstract}

\maketitle
\section{Introduction}
Thermally induced fluctuations and energy dissipation are intimately linked quantities: they both arise from the coupling of a system with its environment. When an unsolicited mechanical system has a high (kinetic, potential) energy for example, it will be damped by its environment, progressively losing this energy to reach a minimum of potential. The energy transfer occurs in the opposite direction if the system is too quiet: random driving from the environment, acting as a thermostat at temperature $T$, induce fluctuations known as thermal noise in the observables of the system. In statistical physics, the equilibrium is defined by  the steady state where on average the energy fluxes cancel out. The amplitude of the thermal noise is then accurately described by the Fluctuation-Dissipation Theorem (FDT), which states that the magnitude of fluctuations is proportional to temperature and dissipation~\cite{Callen1951}. From the FDT, one can accurately describe the thermal noise of any system in equilibrium, such as the Johnson-Nyquist noise in electrical impedances~\cite{Johnson1928,Nyquist1928}, the Brownian motion of particle in a fluid~\cite{Einstein-1905,Perrin-1909}, the mechanical noise of atomic force microscopy (AFM) cantilevers~\cite{Butt1995}, the thermal induced surface waves on a liquid~\cite{Tay-2008}, etc.

In real life, equilibrium is however the exception rather than the rule: living matter, operating devices, and unsteady systems are all out of equilibrium and experience unbalanced energy fluxes with the environment. In many cases, their random fluctuations due to their temperature cannot be described in a universal way. Extensions of the FDT would be useful to understand the thermal noise in such ubiquitous situations. This is especially pertinent for mesoscopic scales, where thermal noise and common deterministic operations have a similar amplitude, or high precision measurements, where any noise is a source of uncertainty that should be avoided or at least characterized. We are interested here in a simple case: a mechanical system subject to a steady heat flux. The situation is pertinent for micro devices whose position is measured with a laser, such as AFM cantilevers~\cite{Meyer1988,Aguilar2015} heated by absorbing a fraction of the light. It is also meaningful for gravitational wave interferometers~\cite{Conti2013,Conti2014}, where a heat flux occurs in the suspension system of the mirrors of the instrument under intense laser radiation. In both cases, thermal noise degrades the performance of the apparatus, and should be minimized. Once thermal fluctuations are understood, they can also be turned into a measurement tool: they can for example help identify dissipation sources, turning the usual annoyance into a useful signal.

In this article, we tackle the thermal noise of microcantilevers subject to a steady heat flux. As in previous works~\cite{Geitner2017,Fontana2020,Fontana2021}, we use an extension of the FDT~\cite{Falasco2014,Geitner2017,Komori2018} to deal with these spatially extended systems presenting a temperature profile, rather than a single temperature corresponding to the thermostat. In these studies, we assumed a dissipation mechanism for elastic energy (located in a single point~\cite{Geitner2017,Fontana2020,Fontana2021}, or uniformly distributed along the cantilever~\cite{Geitner2017}) and demonstrated that the measured thermal noise amplitude was compatible with such hypotheses for damping. As a further insight into this playground, we measure here the dissipation in parallel to the fluctuations on three different samples having distinct damping mechanisms. We show that the dependency of the dissipation and the fluctuations on the external heating is reasonably captured by the model, and concluded that both quantities are indeed linked by the proposed extended FDT.

The article is organized as follows. In section \ref{section:methods}, we first present the methods: samples, measurement device, mechanical modes of the cantilevers, experimental procedure to create the nonequilibrium steady state (NESS), extraction of the temperature field, thermal noise amplitude and global dissipation, and finally the expected link between fluctuation and dissipation. In section \ref{section:results}, we present the measurement results for the three samples, and check the proposed framework. Finally, we briefly discuss the success and shortcomings of the model before concluding in section \ref{section:ccl}. 

\section{Methods} \label{section:methods}
\begin{figure}
\begin{center}
\includegraphics[width=\columnwidth]{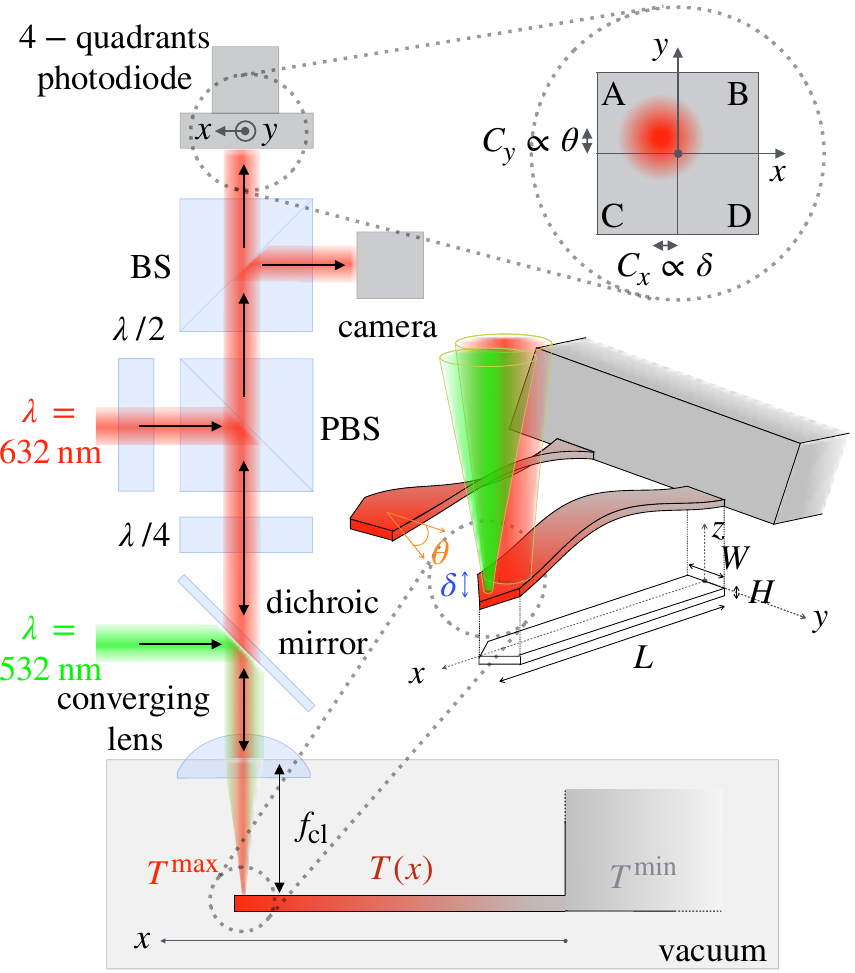}
\caption{Experimental setup: The deflection and torsion of a cantilever are captured thanks to the optical lever technique. The red laser beam ($\SI{1}{mW}$ at $\SI{633}{nm}$), focused on the cantilever tip, is reflected towards the four-quadrant photodiode. This sensor records the temporal signals of deflection $\delta(t)$ and torsion $\theta(t)$. A green laser beam ($0-\SI{12}{mW}$ at $\SI{532}{nm}$) focused close to the tip of the triangular end of the cantilever acts as the heater. A camera is used to visualize the position of both lasers on the sample. The cantilever, in vacuum at $\SI{5e-6}{mbar}$, is monolithically clamped to its macroscopic chip, which is thermalized at room temperature $T^\mn$.}
\label{Fig.setup}
\end{center}
\end{figure}

The experimental setup is depicted in fig.~\ref{Fig.setup}. The physical system under study is a silicon microcantilever, whose thermal fluctuations are measured close to its free end with the optical lever technique~\cite{Jones1961,Meyer1988}. A red laser beam (633 nm) is focused with normal incidence on the cantilever, and its reflection is collected with a four-quadrant photodiode. A green laser (532 nm), focused close to its free end, is partially absorbed and acts as a heat source, creating a temperature profile along the sample. The cantilever, in vacuum at $\SI{5e-6}{mbar}$, is monolithically clamped to its macroscopic chip which is thermalized at room temperature $T^\mn$.
\subsection{Microcantilevers}
The physical system consists of silicon microcantilevers of typical length $L=\SI{500}{\micro m}$, with width $W$ and thickness $H$ which depend on the sample. We discuss here the three samples considered in this work. 

The first sample is a silicon cantilever, $W=\SI{100}{\micro m}$ wide and $H=\SI{1}{\micro m}$ thick (Nanoworld Arrow TL-8~\cite{ARROW}), with a triangular free end such as the one sketched in fig.~\ref{Fig.setup}. This cantilever is of particular interest because it is the same sample studied in the previous NESS experiments in the group, first focusing on the flexural fluctuations~\cite{Geitner2017} and more recently adding the torsional ones~\cite{Fontana2020}. In both these works it is demonstrated how, Brownian fluctuations-wise, the cantilever is almost insensitive to the thermal flux it withstands. In this work we intend to complete the previous work by adding the study of the mechanical dissipation of the sample alongside its thermal fluctuations. We refer to this sample as C100.

The second sample is a silicon cantilever, $W=\SI{30}{\micro m}$ wide and $H=\SI{2.67}{\micro m}$ thick (BudgetSensors AIO-TL~\cite{AIOTL}), also with a triangular tip at its end. This sample, albeit being made purely of silicon as C100, shows a substantially different behavior with respect to C100. Indeed, the system is sensitive to the temperature profile along the system, from both fluctuations and dissipation points of view. We refer to this cantilever as C30.

The third sample is the same C30 cantilever, additionally coated with a Tantala (Ta$_2$O$_5$) thin layer by the Laboratoire des Mat\'eriaux Avanc\'es (LMA, Lyon, France)~\cite{Li2014,PEDURAND2019}. With this sample, we study the effect of a distributed dissipation (due to the coating) on the thermal fluctuations. This cantilever is the same one used in past experiments of the group~\cite{Geitner2017}, where it was shown how the flexural thermal noise is strongly dependent on the temperature profile imposed on the system. In this work, we expand these results for a torsional resonance mode and analyze the behavior of the dissipation. We refer to this sample as C30C.

A summary of the characteristics of the different samples can be found in Table~\ref{tab.samples}. A short discussion on geometrical differences between cantilevers C30 and C100 is given in Appendix~\ref{appendix}.
\begin{table}[htbp]
\caption{Cantilevers studied in this article and their characteristics.}
\begin{center}
\begin{tabular}{ | c | c | c | c | c | } 
\hline
Sample & Length  & Width & Thickness & Ta$_2$O$_5$ coating \\
reference & $L$ [$\SI{}{\micro m}$] & $W$ [$\SI{}{\micro m}$] & $H$ [$\SI{}{\micro m}$] & (each side) [$\SI{}{\mu m}$] \\
\hline
C100 & 500 & 100 & 1 & - \\
\hline
C30 &  500 & 30 & 2.67 & - \\
\hline
C30C &  500 & 30 & 2.67 & 0.3\\
\hline
\end{tabular}
\end{center}
\label{tab.samples}
\end{table}
\subsection{Experimental setup}
As illustrated in Fig.~\ref{Fig.setup}, the red laser ($\SI{1}{mW}$ at $\SI{633}{nm}$) enters the system through a half-wave plate ($\lambda/2$) which tunes its polarization so that after passing through the polarizing beam splitter (PBS) the light is directed towards the cantilever. It then passes through a quarter-wave plate ($\lambda/4$), a dichroic beam splitter, and a converging lens ($f_\cl=\SI{30}{mm}$) which focuses the beam on the cantilever tip. The waist diameter is tuned to roughly $\SI{100}{\micro m}$ to maximize sensitivity~\cite{Gustafsson1994}. The lens is also used as the light port to the vacuum chamber. Light is reflected back on the same path from the cantilever. The second passage through the quarter-wave plate rotates the polarization perpendicular to the initial one, and therefore the return beam passes straight through the PBS. A final beam splitter (BS) divides it towards an optical camera, used to position the lasers on the cantilever, and the four-quadrant photodiode. A motorized 2D translation platform controlling the position of the sensor in these directions is used in the calibration step (see Ref.~\onlinecite{Fontana2020} for details). 

The green laser beam ($0-\SI{12}{mW}$ at $\SI{532}{nm}$) focused close to the tip of the cantilever acts as the heater. It is directed towards the cantilever by the dichroic mirror and through the lens. Part of the intensity is absorbed and creates a heat flux, and another part is reflected and runs through the same path out of the system. The two laser spots do not overlap in order to avoid mutual disturbances. We discuss the temperature of the cantilever under the action of the heater in section \ref{sec.temperature}. 
\subsection{Mechanical resonance modes}\label{sec.displacements}
\begin{figure}
\begin{center}
\includegraphics[width=\columnwidth]{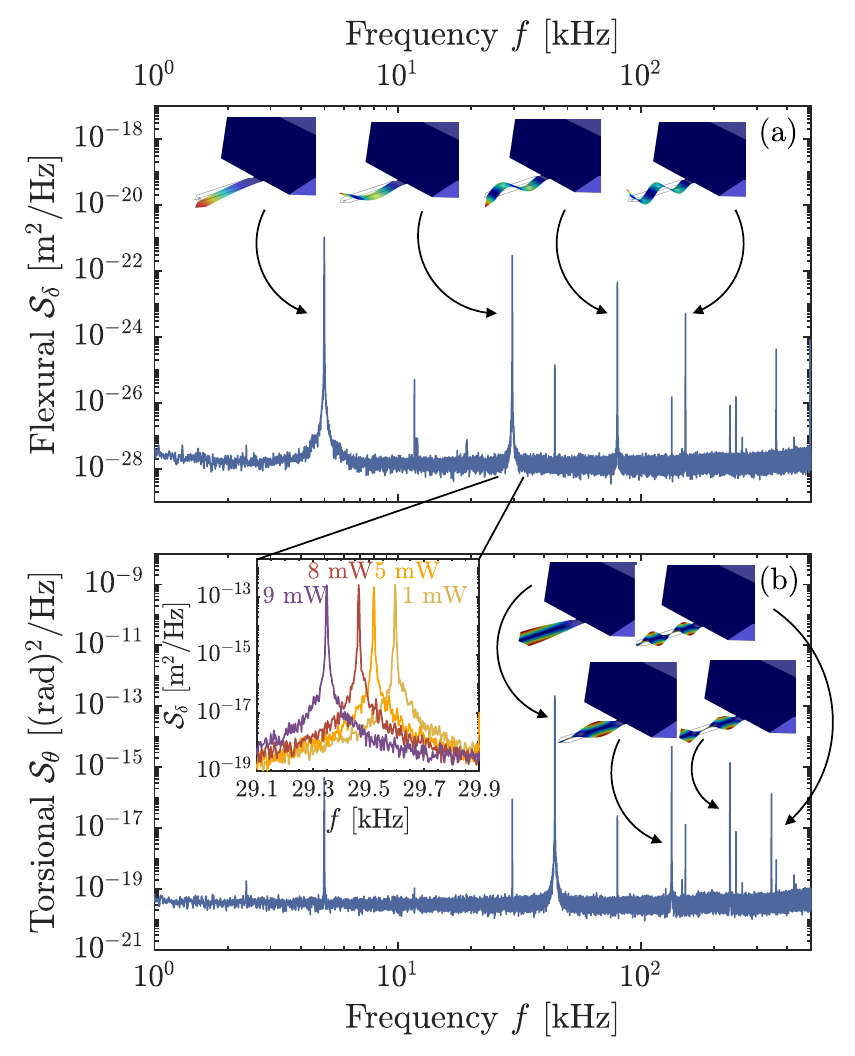}
\caption{PSDs of the (a) thermal-noise-induced deflection and (b) torsion of the cantilever, where each resonance is identified as a sharp peak with a quality factor in the range of tens of thousands. The modes can safely be considered decoupled and each can be treated as a simple harmonic oscillator. In the inset, a zoom-in around the second flexural resonance shows how the resonance is redshifted with the laser power increasing. The shapes of the modes are simulated in COMSOL~\cite{COMSOL} and shown as snippets corresponding to each peak, yielding resonance frequencies very close to the ones found in our experiment and in agreement with the Euler-Bernoulli description.}
\label{Figs.pectrad}
\end{center}
\end{figure}

The photodetector captures four light power signals, which combined give two contrasts $C_x$ and $C_y$ (ratio of the difference over the sum along the $x$ and $y$ axes respectively). These signals are proportional to the angle of the beam upon reflection on the cantilever. The contrast $C_x$ leads to the calibrated flexural angle $\vartheta$ (in radians), which can be converted to the deflection $\delta$ (in meters), while the contrast $C_y$ is proportional to the torsional angle $\theta$ (in radians). The conversion factors and the calibration are carefully discussed in Ref.~\onlinecite{Fontana2020}. Computing the Power Spectrum Density (PSD), we identify the normal modes of the cantilever, which are shown in fig.~\ref{Figs.pectrad}. The spectra are shot-noise limited and the thermal noise-driven resonance peaks have a high signal-to-noise ratio.

Typical measurements allow us to explore a wide range of frequencies, where the observable number of modes depends on the geometry of the sample. In the case of C100, this is up to 11 flexural and 8 torsional modes; for the C30 and C30C cantilevers we can detect up to 5 flexural and 1 torsional mode. In order to ensure we correctly identify the resonances, we simulate the cantilever's eigenmodes in COMSOL~\cite{COMSOL}. Indeed, due to the imperfect orientation of the photodetector, torsional signals are visible in the flexural PSD and vice versa (see fig.~\ref{Figs.pectrad}), and the simulation helps us qualitatively distinguish the two motions, especially at high frequency where amplitudes are intrinsically small and vanish close to nodes. Another important contribution of these simulations is to prove we can access all the resonances in the available frequency range: this is indeed true except for one lateral mode (oscillations in the $x-y$ plane), undetectable with our setup. 

Due to experimental constraints, in this article, some modes are excluded from the analysis. Flexural mode 1 is often discarded because of self-oscillations~\cite{Metzger2008,Fontana2020}, while some modes can be undetectable due to the probing point being close to a node of sensitivity~\cite{Fontana2020}. 
\subsection{Experimental procedure}\label{sec.proc}
In order to probe the thermal noise of the cantilever in a NESS, we increase the temperature at the tip of the sample through the increase of the injected power $P$. We then perform a power ramp, going from 0 to $\SI{12}{mW}$ and back, in order to compare the results at increasing and decreasing temperature. At each power we record the Brownian motion of the cantilever. In order to reduce the statistical uncertainty on the measured noise, at each power step we record a large number $N^\mathrm{meas}$ of temporal signals (between 40 and 75), each $t^\meas = \SI{2}{s}$ long and sampled at $\SI{2.5}{MHz}$.
\subsection{Temperature}\label{sec.temperature}
The absorbed power of the green laser, placed at the tip of the cantilever, creates a heat flux along the length of the sample. This generates a temperature profile $T(x)$ between the maximal temperature at the tip of the sample $T^\mx$ and the temperature at the base $T^\mn = \SI{295}{K}$, which is kept constant by the contact with the macroscopic chip acting as the thermal reservoir. On the other side, $T^\mx$ and $T(x)$ vary with the absorbed power according to Fourier's law. The average temperature $T^\avg = \int_0^L \d x \, T(x)/L$ is used to characterize the the nonequilibrium state of the cantilever: the measured amplitude of the thermal fluctuations in a NESS at $T^\avg$ is compared to the one the system should have if it was in equilibrium at the same $T^\avg$. Its estimation is thus paramount. 

As shown in fig.~\ref{Fig.setup}, the flexural resonance frequencies $f_n$ are sensitive to the temperature changes in the cantilever, mostly through the variations of the Young modulus of silicon $Y$. While this is discussed in detail in Ref.~\onlinecite{Pottier-2021-JAP129}, we briefly recall it here. In a first approximation, the frequency shift $\Delta f_n = f_n - f_n^\mn$ can be modeled as:
\begin{equation}
\label{eq.dffgen}
\frac{\Delta f_n}{f^\mn_n} = \frac{1}{2} \frac{\int_0^L\d x\, \Delta Y (T(x)) \psi_n(x)^2}{Y^\mn \int_0^L\d x\, \psi_n(x)^2} = g_n(T),
\end{equation}
where $\Delta Y = Y - Y^\mn$ and the superscript $\mn$ stands for the reference value of the quantity at $T^\mn$. The function $\psi_n(x)=\phi^{\prime \prime}_n(x)$ is the curvature of the normal mode $\phi_n$ considered. The functions $g_n$ are governed by the temperature dependency of the Young modulus, which for silicon is tabulated~\cite{Gysin2004}. In the case of C30C, the cantilever is the same sample used in ref.~\onlinecite{Geitner2017}, where a calibration of $g_n$ was performed. Finally, in Ref.~\onlinecite{Pottier-2021-JAP129} it is shown how inverting the functions $g_n$ yields access to the average temperature of the cantilever at each measured frequency shift. Therefore, for all the thermal noise measurement presented in this work we can associate a $T^\avg$ at each imposed heating power. 

The estimation of the uncertainty on $T^\avg$ is discussed in Refs.~\onlinecite{Pottier-2021-JAP129,Fontana2020,Fontana2021}. In a nutshell, for each of the $N^\mathrm{meas}\sim50$ time recordings at a specific power, we retrieve with a fit of the thermal noise peak the value of $f_n$ for each mode $n$. From the uncertainty on the fit parameter, the dispersion between the modes, the statistical uncertainty computed on the $N^\mathrm{meas}$ recordings (computed as their standard deviation over $\sqrt{N^\mathrm{meas}-1}$), the uncertainty on the function $g_n$, and standard error propagation rules, we deduce the uncertainty on $T^\avg$.
 
\subsection{Thermal noise}
When the heating power is zero, the cantilever is considered in thermal equilibrium with the environment at a temperature $T^\mn$. As we can see from fig.~\ref{Figs.pectrad}, the resonances are well separated in frequency, have a high signal-to-noise ratio, and a quality factor larger than $1000$. Hence, we can model each flexural and torsional mode as an independent oscillator, and thus we can apply the equipartition principle to each resonance:
\begin{equation}
\label{eq.EP}
k_n \langle \delta_n^2 \rangle_\EQ = \kappa_m \langle \theta_m^2 \rangle_\EQ = k_\B T^\mn,
\end{equation}
where $k_\B$ is the Boltzmann's constant, and
\begin{equation}
\label{eq.stiff}
\begin{split}
k_n = m^\eff \omega_n^2, \quad \kappa_m= J^\eff \omega_m^2, 
\end{split}
\end{equation}
are the flexural (index $n$) and torsional (index $m$) stiffnesses respectively, with $J^\eff = m^\eff W^2/3$ the inertial moment of the beam, $m^\eff$ its effective mass and $\omega_{n,m} = 2 \pi f_{n,m}$. The subscript $_\EQ$ emphasize here that the system is considered in equilibrium (implying that we neglect the absorption of the red laser beam). The quantities $\langle \delta_n^2 \rangle, \langle \theta_m^2 \rangle$ are the thermal fluctuations, calculated as the area under the resonance peak once the background noise contribution is subtracted~\cite{Fontana2020}. 

When the cantilever is in a NESS due to the presence of the heat flux, we extend Eq.~\eqref{eq.EP} to define a fluctuation temperature $T^\fluc$ as 
\begin{equation}
\label{eq.Tfluc}
\begin{split}
T^\fluc_n & \equiv \frac{k_{n} \langle \delta^2_n \rangle}{k_B} = \left(\frac{f_n}{f_n^\mn}\right)^2\frac{\langle \delta^2_n \rangle_\NESS}{\langle \delta^2_n \rangle_\EQ} T^\mn, \\
T^\fluc_m & \equiv \frac{\kappa_{m} \langle \theta^2_m \rangle}{k_B} = \left(\frac{f_m}{f_m^\mn}\right)^2\frac{\langle \theta^2_m \rangle_\NESS}{\langle \theta^2_m \rangle_\EQ} T^\mn.
\end{split}
\end{equation}
This quantity represents the temperature the system fluctuates at when it is in an out-of-equilibrium state. Indeed, in this condition, no thermodynamic temperature can be defined and the equipartition principle cannot be applied. Nevertheless, the amplitude of the fluctuations and the resonance frequency can be measured; consequently a temperature $T^\fluc$ can be defined. While a single temperature $T$ can be found for all the modes in equilibrium (Eq.~\eqref{eq.EP}), in a NESS $T^\fluc$ is in principle mode dependent since each resonance mode represents in this case a different thermometer. 

The evaluation of the uncertainty on $T^\fluc_{n,m}$ is discussed in Refs.~\onlinecite{Nor2010,Fontana2020,Fontana2021}. Its statistical part is computed from the dispersion around the mean of the $N^\mathrm{meas}$ measurements (as std/$\sqrt{N^\mathrm{meas}-1}$). This uncertainty is intimately linked to the number of independent samples that we extract from one measurement. The relaxation time of one mode (the time it takes to forget its initial conditions) is $\tau_n = 2 /(\varphi_n f_n)$, with $\varphi_n$ the dissipation associated to the mode. For a recording time $t^\mathrm{meas}$, we have $N^\mathrm{indep}=t^\mathrm{meas}/\tau_n$ independent samples. Both $\varphi_n$ and $\omega_n$ are increasing with the mode number $n$, and thus the statistical uncertainty decreases with $n$. For large $n$ however, the signal-to-noise ratio (ratio of thermal noise to floor noise) decreases and the uncertainty rises again: intermediate $n$ have the lowest statistical uncertainty. Apart from this unavoidable contribution, we must consider another source of error: during the measurement, the laser position can slowly shift, mainly due to experimental drifts. This effect causes a change in the sensitivity of the experiment; therefore, we ascribe it to a systematic uncertainty. As it turns out, this can be the main contribution depending on the mode. Indeed, the amplitude of the change in sensitivity is all the more important when the measurement point is closer to a node of the mode. Since the laser is focused close to the free end of the cantilever, the first resonances are less affected. The statistical and systematic uncertainties of $\Tfluc_{n,m}$ are finally quadratically summed.

The right-hand-side definition of $T^\fluc$ in Eq.~\eqref{eq.Tfluc} has the advantage of yielding a simple calculation of $T^\fluc$ as the ratio of measured quantities: the amplitude of the nonequilibrium fluctuations ($\NESS$) and the equilibrium ones (EQ). Recalling the discussion in section \ref{sec.proc}, this corresponds to normalizing each nonequilibrium measurement by the average of the first and last recordings on the power ramp. Finally, the prefactor with the ratio of the resonance frequencies in Eq.~\eqref{eq.Tfluc} takes into account the changes in stiffness with the temperature (Eq.~\eqref{eq.stiff}), since $m^\eff$ and $J^\eff$ are supposed constant.

The nature of the fluctuation temperature is related to the existence of a nonequilibrium Fluctuation-Dissipation Theorem. Carefully extending this relation, which is normally valid solely in equilibrium, for a system with a temperature profile $T(x)$, it is possible to show that $T^\fluc$ is related to the normalized local mechanical energy dissipation $w^\diss(x)$ in the cantilever and the temperature profile as~\cite{Geitner2017,Komori2018,Fontana2020}
\begin{equation}
\label{eq.EPNESS}
T^\fluc_{n,m} = \int_0^L \d x  \, T(x) w^\diss_{n,m}(x).
\end{equation}
This relation tells us that the amplitude of the fluctuations is then related to the temperature profile weighed by the dissipation profile, i.e. the locations with the higher dissipation contribute more to the total thermal noise. We discuss the nature of the damping in the system in the next section.
\subsection{Dissipation}

In the experiment, the damping of the system is measured through a fit of the PSD around the resonance frequency using the following expression, for example for a flexural mode $n$:
\begin{equation}
\label{eq.fit}
\mathcal{S}_{\delta_n}(f) = \frac{2 k_\B T}{\pi m^\eff \omega_n^2 f} \frac{f_n^4 \varphi_n}{(f^2-f_n^2)^2 + (f_n^2 \varphi_n)^2}.
\end{equation}
The loss angles $\varphi_{n,m}$ are extracted from the fits of the $N^\mathrm{meas}$ spectra corresponding to the recordings, and we compute their expectation value (as the mean) and statistical uncertainty (as std/$\sqrt{N^\mathrm{meas}-1}$). $\varphi_{n,m}$ represent the global dissipation of the cantilever for each mode. It can have the various microscopic origins, and can depend slowly on frequency. However for each mode, strongly peaked around resonance, it corresponds to the inverse of the quality factor $Q$, and we have no access to its value out of the resonance frequency.

The cantilever being held in vacuum, the dissipation of the bulk material (silicon) arises from two mechanisms: clamping losses~\cite{Hao2003} and internal damping~\cite{Paolino2009} (also called viscoelasticity). The latter can be due to the presence of defects in the cantilever or thermoelasticity~\cite{Nowick1972}. The presence of the coating adds damping via its own internal damping and losses at the interface with the substrate~\cite{Granata2020}. In the end, the effect of these processes is the loss angle $\varphi$. In principle, $\varphi$ depends on the frequency $f$, the temperature $T$, the presence of defects, and consequently the spatial coordinate $x$. The loss angle can be thought of as the imaginary part of the static stiffness:
\begin{equation}
\begin{split}
k & = \frac{3 I}{L^3} Y= k^0(1+i \varphi_Y), \\
\kappa & = \frac{4 I}{L} S = \kappa^0(1+i \varphi_S), \\
\end{split}
\end{equation}
with $I=WH^3/12$ the second moment of area of the sample, and $S$ the shear modulus. Here the flexural and torsional loss angles,  $\varphi_Y$ and $\varphi_S$ respectively, are mainly due to the imaginary part of the Young modulus and shear modulus. 

The loss angles $\varphi_{n,m}$ of the modes are expressed through~\cite{Geitner2017,Komori2018,Fontana2020}:
\begin{equation}
\label{eq.locdiss}
\begin{split}
\varphi_n\{T(x)\} & = \int_0^L \d x \, \varphi_Y(x,f_n,T(x)) \psi_n(x)^2, \\
\varphi_m\{T(x)\} & = \int_0^L \d x \, \varphi_S(x,f_m,T(x)) \psi_m(x)^2,
\end{split}
\end{equation}
with $\psi_{n,m}(x)$ the local curvature: $\psi_n(x)=\phi_n^{\prime\prime}(x)$ for the flexural normal mode, where $\phi_n(x)$ is the local deflection, and $\psi_m(x)=\phi_m^{\prime}(x)$ for the torsional normal mode, where $\phi_m(x)$ is the local transverse slope. Using these notations, the two expressions in Eqs.~\eqref{eq.locdiss} are equivalent for flexural and torsional modes simply interchanging subscripts $n$ by $m$ and $Y$ by $S$, thus in the following we display only the equations for flexural modes, but everything applies directly to torsional modes as well. The spatial profile $\varphi_{Y,S}$ is in general not experimentally accessible, and thus neither is the normalized local dissipation $w^\diss_{n,m}(x)$, defined as~\cite{Geitner2017,Komori2018,Fontana2020}:
\begin{equation}
\label{eq.wdiss}
w^\diss_n(x) = \frac{1}{\varphi_n\{T(x)\}} \varphi_Y(x,f_n,T(x)) \psi_n(x)^2.
\end{equation}
Nevertheless, in section \ref{subsection.local-global} below, we show that, if certain hypotheses regarding the temperature profile and the local damping are satisfied, $w^\diss$ becomes experimentally accessible. In this case, the fluctuation temperature $T^\fluc$ can be theoretically calculated through Eq.~\eqref{eq.EPNESS} and compared to the experimental results. 

\subsection{Local properties vs global measurements}\label{subsection.local-global}

The elastic properties of the silicon and of the optional coating depend only weakly on the temperature: the resonance frequencies of the normal modes change for instance in the per thousand range when the average temperature doubles. We make the hypothesis that the dissipative part of the elastic moduli changes accordingly, so that a second order expansion of $\varphi_Y$ in $T$ is enough on the explored temperature range. If the damping is distributed, we also make the hypothesis that the material properties are uniform (independent of the position), so that
\begin{equation}
\label{eq.varphiquadDT}
\varphi_{Y} (x,f_{n},T) = \alpha_{n} +   \beta_{n}\Delta T(x) + \gamma_{n} \Delta T (x)^2.
\end{equation}
The weak dependency of properties on frequency is captured by the mode number dependency, sampling properties only around the resonance frequency $f_{n}$. Let us finally suppose that the temperature profile is linear: $\Delta T(x) = \Delta T^\mx x/L$. This assumption is equivalent to considering that the thermal conductivity of the cantilever is independent of temperature. It is reasonable at the considered heating powers for samples C30 and C30C, and it was verified thanks to numerical simulations and in previous experiments~\cite{Pottier-2021-JAP129}. In such a case, Eq.~\eqref{eq.varphiquadDT} is written
\begin{equation}
\label{eq.varphiquadTmax}
\varphi_{Y} (x,f_{n},T) = \alpha_{n} +   \beta_{n}\Delta T^\mx \frac{x}{L} +  \gamma_{n}\left(\Delta T^\mx \frac{x}{L}\right)^2.
\end{equation}
Injecting this expression in Eqs.~\eqref{eq.locdiss}, we compute the global dissipation of each mode as
\begin{equation}
\varphi_{n} = a_{n} + b_{n} \Delta T^\mx + c_{n} (\Delta T^\mx)^2,
\end{equation}
where the set of coefficients $a_{n}$, $b_{n}$ and $c_{n}$ is directly linked to the expansion coefficients of the elastic moduli $\alpha_{n}$, $\beta_{n}$ and $\gamma_{n}$:
\begin{equation}
\label{eq.abcvsalphabetagamma}
\begin{split}
a_{n} & = \frac{\alpha_n}{L}\int_0^L \d x \, \psi_{n}(x)^2, \\
b_{n} & = \frac{\beta_n}{L}\int_0^L \d x \, \frac{x}{L} \psi_{n}(x)^2,  \\
c_{n} & = \frac{\gamma_n}{L}\int_0^L \d x \, \frac{x^2}{L^2} \psi_{n}(x)^2. 
\end{split}
\end{equation}

Experimentally, we can measure the dissipation $\varphi_n$ as a function of $\Delta T^\mx$, thus when a parabolic fit is pertinent to describe these data, we can extract coefficients $\alpha_{n}$, $\beta_{n}$ and $\gamma_{n}$ from the fit parameters and Eqs.~\eqref{eq.abcvsalphabetagamma}. We therefore can compute the local loss angle (Eq.~\eqref{eq.varphiquadTmax}), then the normalized dissipation (Eq.~\eqref{eq.wdiss}), and finally the expected amplitude of the fluctuations (Eq.~\eqref{eq.EPNESS}). We refer to this computed temperature as theoretical to distinguish it from the measured one, though it is expressed from experimentally accessible parameters:
\begin{align}
\label{eq.ttheo}
T^\theo_{n}  = & \, T^\mn + \frac{1}{\varphi_{n}}\frac{1}{L} \int_0^L \d x \, \Delta T^\mx \frac{x}{L} \psi_{n}^2(x) \\
& \left(\alpha_{n} +   \beta_{n}\Delta T^\mx \frac{x}{L} +  \gamma_{n}\left(\Delta T^\mx \frac{x}{L}\right)^2\right).  \nonumber
\end{align}
In the case of a distributed damping, we thus have a strategy to assess the validity of our model; ie., compute the expected value of the fluctuation from the measurement of dissipation.

\section{Results}\label{section:results}
In this section we discuss the results of the measurements on the three cantilevers. In each section we show the measured fluctuation temperatures $T^\fluc_{n,m}$ alongside the estimated damping $\varphi_{n,m}$, and we link them through Eq.~\eqref{eq.EPNESS}. 
\subsection{C100}
\begin{figure}[t]
\includegraphics[width=\columnwidth]{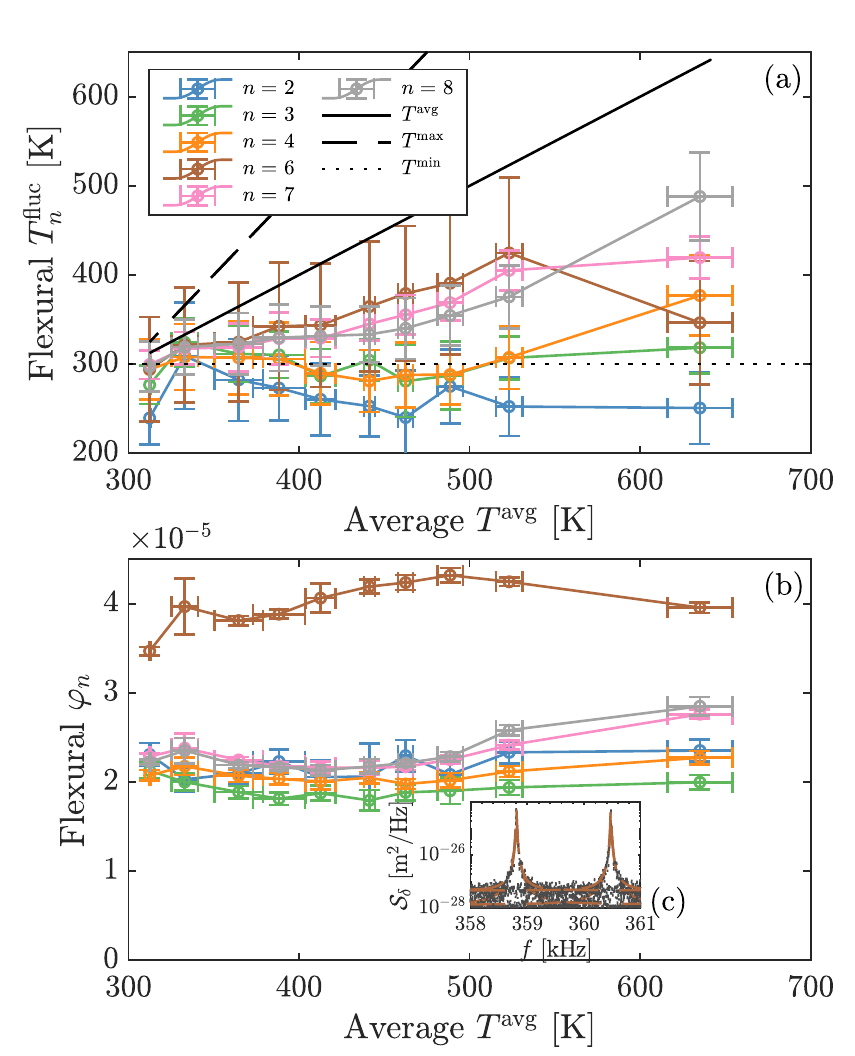}
\caption{Cantilever C100, flexural modes. (a) The thermal noise amplitude $T^{\fluc}_n$ is shown with respect to $T^\avg$. The black solid line represents the equilibrium temperature, i.e. the fluctuations an object would show had it been in thermal equilibrium with a thermal bath at $T^\avg$. All the modes lie below this line, showing a dearth of thermal noise. Furthermore, we note how they are also much lower than the maximal temperature of the system, represented by the black dashed line. The modes shown span from 2 to 8, excluding mode 5 because of the laser probe being on a mechanical node. (b) The measured loss angles $\varphi_n$ show little change with the average temperature. We note also how the loss angle of the mode $n=6$ is roughly twice that of the other modes. (c) The PSD of the same mode at two different powers alongside the fit with Eq.~\eqref{eq.fit}.}
\label{fig.C100.D}
\end{figure}
\begin{figure}[t]
\includegraphics[width=\columnwidth]{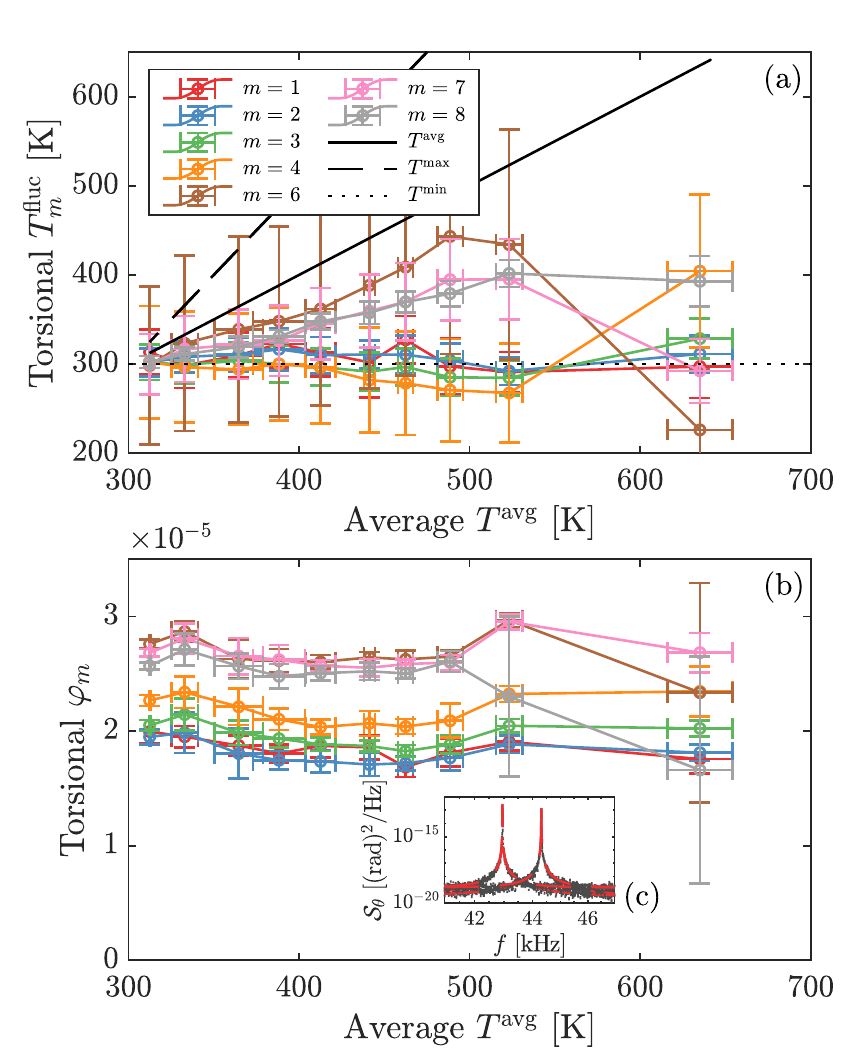}
\caption{Cantilever C100, torsional modes. (a) The thermal noise content of the first eight torsional modes (except the fifth due to the sensitivity being too low) is shown to be roughly independent of the temperature profile imposed on the system. (b) The loss angle also shows little changes at different temperatures. (c) The resonance $m=1$ at two powers alongside the fit with Eq.~\eqref{eq.fit}.}
\label{fig.C100.T}
\end{figure}
Thermal noise measurements of the C100 sample are shown in Figs. \ref{fig.C100.D}(a) and \ref{fig.C100.T}(a), which as mentioned are the same presented in Ref.~\onlinecite{Fontana2020}. We note how, while the temperature $T^\avg$ increases, the thermal fluctuations of the cantilever are roughly unchanged for both flexural and torsional modes. In order to interpret this dearth of fluctuations through Eq.~\eqref{eq.EPNESS}, we display the evolution of the loss angles $\varphi_{n,m}$ in Figs.~\ref{fig.C100.D}(b) and \ref{fig.C100.T}(b). One can understand the lack of fluctuations assuming that the C100 sample is dominated by clamping losses, i.e. $T^\fluc \approx T^\mn = T(0)$ because $w^\diss(x) \approx w^\diss_0 \delta^D(x)$, where $\delta^D(x)$ is Dirac's distribution. This claim can be assessed with the observation of the damping of the system, which shows little if no evolution at increasing temperature. Two explanations are possible for this phenomenon: either the dissipation is distributed all over the cantilever but it is independent of the temperature, or it is located at a point at constant temperature. The former is unlikely and would not account for the flat behavior of $T^\fluc$: applying Eq.~\eqref{eq.ttheo}, the prediction is a temperature-dependent $T^\fluc$ (linear in $T^\avg$) . The hypothesis of a clamped-based dissipation must thus be considered. In this regard, we may look at two phenomena: clamping shear stresses and clamping-located defects. A simple model for the former~\cite{Hao2003} predicts typical quality factors at least ten times higher than the measured ones; thus it is unlikely to be the dominant phenomenon for this sample. The presence of defects may then be the key. Indeed, the C100 is chemically etched from a single crystal silicon wafer, i.e.\ in principle devoid of internal defects, and the vacuum removes most of the hydrodynamical damping. An imperfect etching at the clamping may still be present, thus lowering the quality factor to the observed values. 

Regardless of its origin, a Dirac delta-type dissipation function $w^\diss(x) \approx \delta^\D(x)$ explains then both the thermal noise and the dissipation measurements through the extended FDT expressed in Eq.~\eqref{eq.EPNESS}. 

\subsection{C30}
\begin{figure}
\begin{center}
\includegraphics[width=\columnwidth]{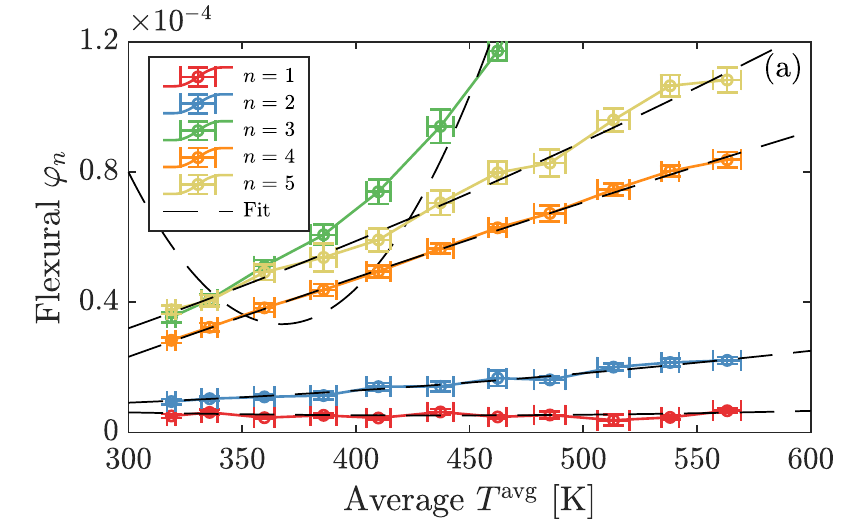}

\includegraphics[width=\columnwidth]{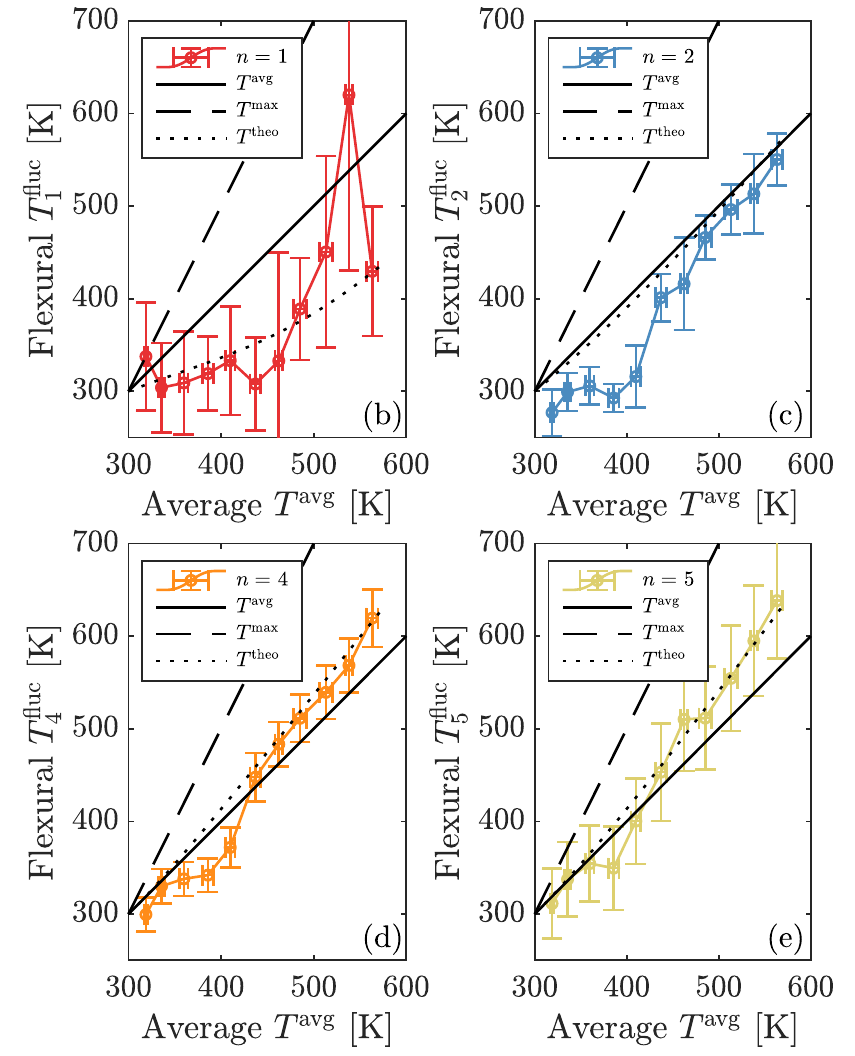}
\end{center}
\caption{Cantilever C30, flexural modes. (a) Dissipation $\varphi_n$ versus $T^\avg$ for flexural modes $n=1-5$, alongside a quadratic fit of each. (b)-(e) Thermal noise amplitude $T^\fluc_n$ alongside its theoretical prediction (dotted line) from Eq.~\eqref{eq.EPNESS} versus $T^\avg$ for the same modes $n$ (excluding $n=3$, shown in fig.~\ref{fig.C30.n3}) . The model nicely predicts the thermal noise evolution, except for a mode 2 and 4 around $T=\SI{400}{K}$ where fluctuations are somewhat below the expected value.}
\label{fig.C30.D}
\end{figure}
\begin{figure}
\includegraphics[width=\columnwidth]{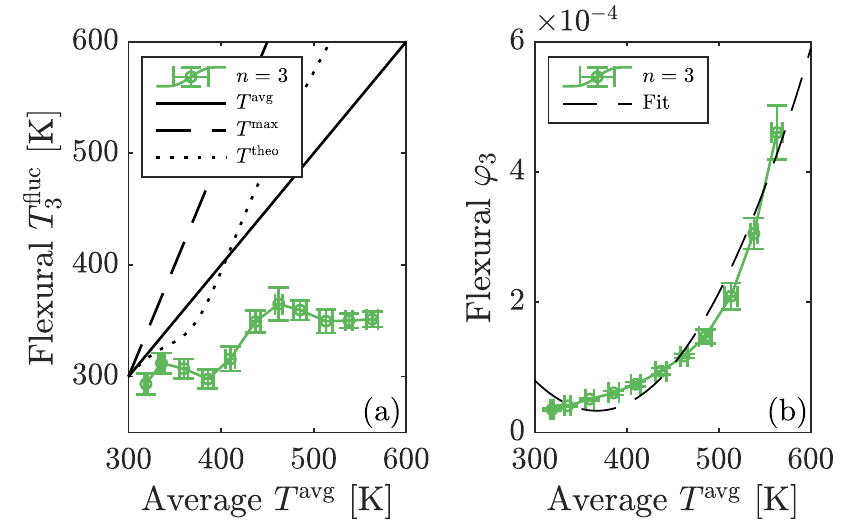}
\caption{Cantilever C30, flexural mode $n=3$. (a) The thermal noise of the third flexural mode shows little to no dependence on the temperature profile, which is the opposite behavior expected looking at the dissipation (b). In this case, the simple quadratic approximation of the dissipation fails, and the model cannot predict the thermal fluctuations.}
\label{fig.C30.n3}
\end{figure}

\begin{figure}
\includegraphics[width=\columnwidth]{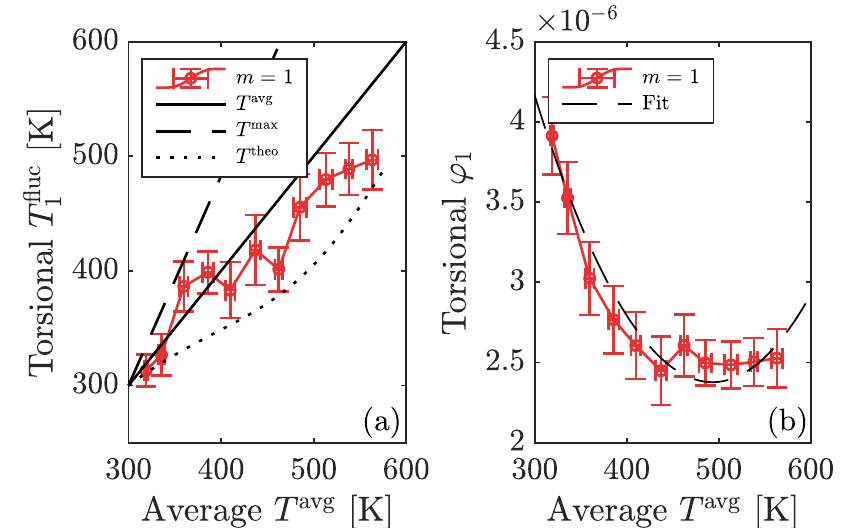}
\caption{Cantilever C30, torsional mode $m=1$. (a) The thermal noise amplitude of the first torsional mode stays roughly below the average temperature line. In this case, the model qualitatively predicts this behavior. (b) The evolution of the dissipation with the temperature and its fit with a quadratic function.}
\label{fig.C30.T}
\end{figure}

The results for the C30 sample are shown in Figs. \ref{fig.C30.D}-\ref{fig.C30.T}. We depict five flexural resonances and one torsional one, with the third resonance mode depicted separately in Fig.~\ref{fig.C30.n3}. We see that apart from the latter, the fluctuation temperature increases with the average temperature, thus showing how we cannot suppose C30 to be a clamping losses-dominated system as C100. The observation of the loss angles confirms this view, since also the damping changes with the temperature. In this case, we can suppose that if clamping losses exist, a distributed damping exists too as it becomes the main contribution when we increase the temperature for the cantilever. Though samples C30 and C100 are made of the same material (silicon), they have slightly different geometries and their manufacturers (and thus manufacturing processes) are different, resulting in different mechanical damping behaviors.

To a reasonable approximation, we observe that the global losses $\varphi_{n,m}$ are a smooth function of the average temperature, and thus of $T^\mx$, so that we can approximate them by a quadratic fit. We can therefore apply the recipe of section \ref{subsection.local-global} to predict the amplitude of the fluctuations. Those are reported as dotted lines in Figs. \ref{fig.C30.D}-\ref{fig.C30.T}. We note how for most modes the overlap of the experimental $T^\fluc$ and this simple model shows an overall reasonable agreement, which suggests that the simple extension of the FDT appears to hold in this distributed-dissipation case. We nevertheless remark how at around $T=\SI{400}{K}$ the thermal content (mostly for mode $n=2$) does not follow exactly the theoretical prediction, and how for the torsion the prediction is a little below the experimental data. Finally, for flexural mode $n=3$ (Fig.~\ref{fig.C30.n3}), the damping and the fluctuations do not draw the same picture: indeed, while the thermal noise is roughly independent of the damping, the latter increases strongly with the temperature. In this case, the simple approximations do not hold. This mode also stand apart from the others on the magnitude of dissipation, which is ten times larger than one would expect from the extrapolation of the smooth (both in temperature and mode number) behavior of modes $1$ through $5$. This odd behavior of these modes probably hints at a different mechanism implied at this specific frequency or mode shape for this cantilever, a mechanism that we could not identify and deemed not representative of the overall behavior.

\subsection{C30C}
\begin{figure}
\includegraphics[width=\columnwidth]{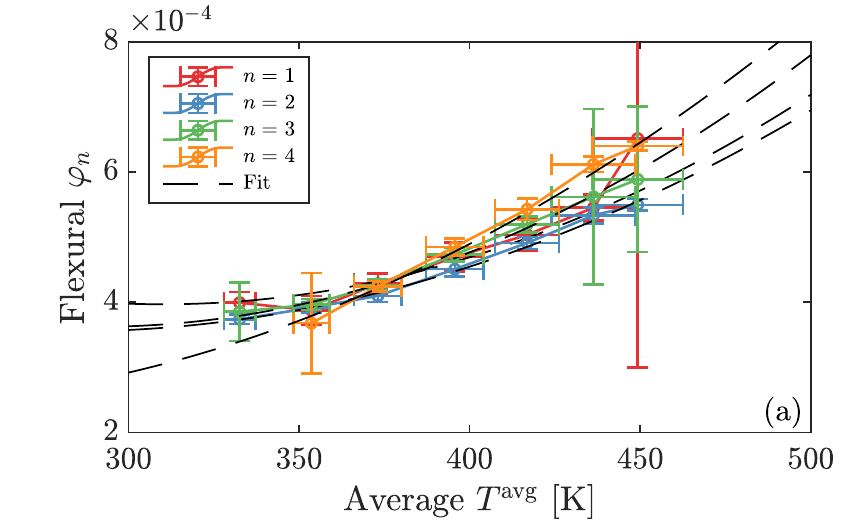}

\includegraphics[width=\columnwidth]{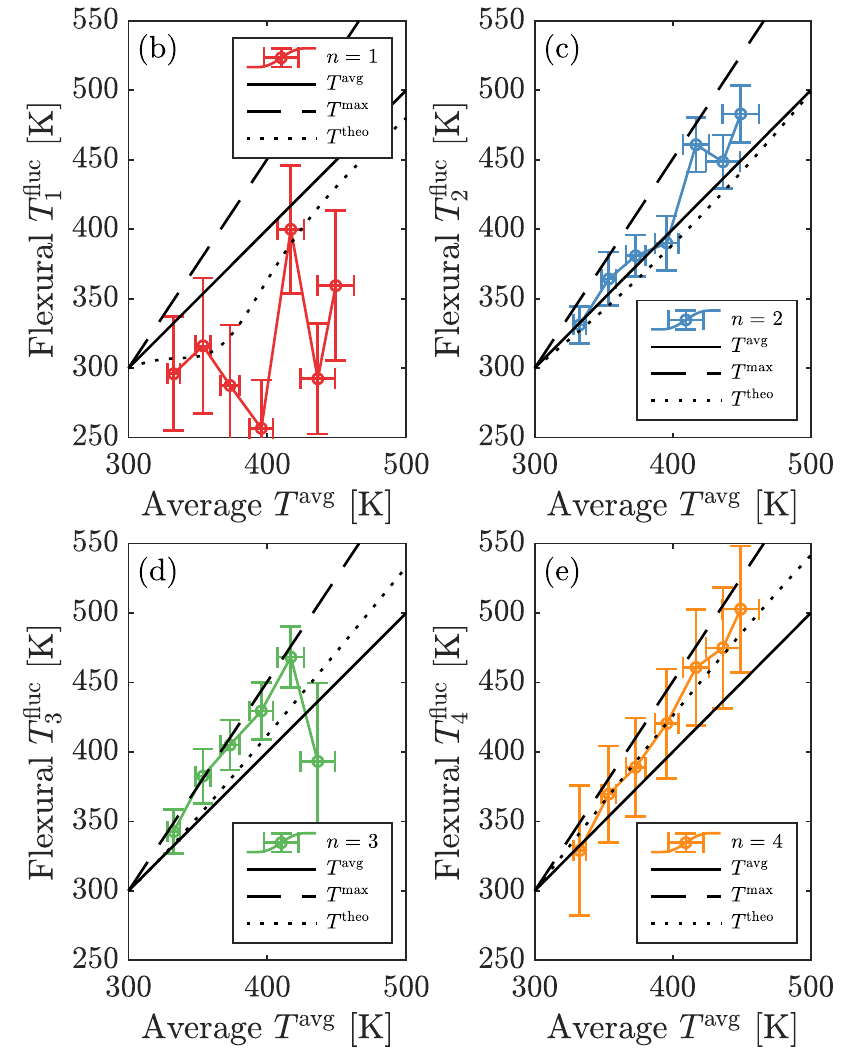}
\caption{Cantilever C30C, flexural modes $n=1-4$. (a) Dissipation $\varphi_n$ versus $T^\avg$, alongside a quadratic fit of each. We note how the coating increases the magnitude of the dissipation at least ten times with respect to the bare C30 sample. (b)-(e) Thermal noise amplitude $T^\fluc_n$ alongside its theoretical prediction (dotted line) from Eq.~\eqref{eq.EPNESS} versus $T^\avg$. The theoretical framework and the simple hypotheses formulated for the C30 sample allow us to approximately predict the nonequilibrium thermal content of the C30C sample. We nevertheless note how mode 3 deviates slightly from the prediction.}
\label{fig.C30C.D}
\end{figure}
\begin{figure}
\includegraphics[width=\columnwidth]{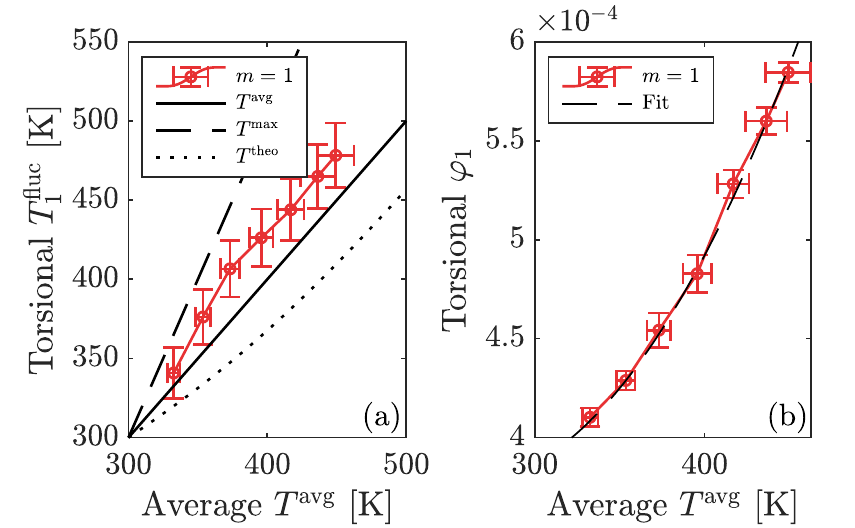}
\caption{Cantilever C30C, torsional mode $m=1$. (a) The theoretical prediction of the thermal noise thanks to the evolution of the dissipation ((b), quadratic fit) would predict an increase of the fluctuations below the average temperature of the system. This is not observed in the measurement. A possible explanation is discussed in the text.}
\label{fig.C30C.T}
\end{figure}
If the C30 cantilever data hint at a distributed dissipation, we expect this phenomenon to be even more striking for the C30C case. Indeed, the coating on this sample adds an additional distributed damping due to the dissipation in the coating thickness~\cite{Li2014}. We can observe this phenomenon comparing Fig.~\ref{fig.C30C.D}(a) with Fig.~\ref{fig.C30.D}(a), where an increase in damping of a factor of ten is measured, alongside the squeezing of all the loss factors alongside roughly the same curve. From the thermal noise point of view, the fluctuation temperatures increase similarly to the C30 case, and the theoretical prediction once again allows us to qualitatively describe them alongside the damping through the simple hypotheses exposed in section \ref{subsection.local-global}. It is to be noted that in this case mode $n=3$ shows higher thermal content than predicted, at the edge of the error bars. This cannot be said for the torsional mode, presented in Fig.~\ref{fig.C30C.T}, for which the damping would predict a fluctuation temperature always below the average one, while the measurement shows otherwise. A possible explanation lies on the uniform hypothesis for the loss angle $\varphi_S(x,\ldots)$. Indeed, a higher value of dissipation close to the free end would stand apart from the hypothesis we used, and raise the theoretical prediction $T_m^\theo$. Even if dissipation processes for the shear modulus and the Young one can be different, a higher dissipation close to the cantilever end would also be compatible with most flexural mode being slightly above the prediction, though mostly within error bars.
\section{Discussion and conclusion} \label{section:ccl}
The comparison of the simultaneous measurement of the nonequilibrium thermal fluctuations and the damping on three different samples allows us to test the simple theoretical prediction represented by the extended equipartition of Eq.~\eqref{eq.EPNESS}. Indeed, this equation simply states that the amplitude of thermal fluctuations of a spatially extended system, when a temperature profile is established, depends on where and at which temperature the dissipation occurs. First, we show how this theoretical framework explains the observed thermal noise and damping on a clamping-located dissipation cantilever. Indeed, the thermal fluctuations and the dissipation are roughly independent of the temperature profile imposed on the system. On different samples (different manufacturer, and optional coating layer), a quite different behavior is observed: the thermal noise increases with the average temperature of the system, as does the dissipation (in general, with the exception of the first torsional mode of cantilever C30). In these cases, the system is dominated by other sources of dissipation, and indeed present loss angles at least twice larger than those of cantilever C100. This can be due to a higher degree of impurity in the system, which create higher viscoelastic losses, or a result of the damping due to the coating. We propose a simple model with a linear dependance of temperature on space and second order expansion of the local dissipation on temperature. Thanks to these simple but reasonable hypotheses, we can then \emph{predict} the fluctuation temperature of the cantilever by just characterizing the damping. Though not always quantitative, the global picture is consistent and predictions of the thermal fluctuations are effective for most of the resonance modes.

Our purpose was to demonstrate the compatibility of the extended equipartition, including the variation of the dissipation with the local temperature field, with the measurement. Though not perfect, this agreement is reasonable. This study sheds light on the power of a simple extension of the FDT for systems far from equilibrium. It expands the previous studies on such systems~\cite{Geitner2017,Fontana2020, Fontana2021} for various damping sources and it includes a theoretical prediction of the fluctuation based solely on the evolution of the dissipation with temperature, whether constant (clamped based damping) or not (distributed damping). Obviously, one could encounter situations where dissipation is a combination of several mechanism with comparable magnitude, and we would then need to mix hypotheses of local (including not only localized at the clamp) and global dissipation. Deviation from the present model could be useful to identify the physical origin or location of the dissipation, with the possibility to use the complementary information from several modes to pin defects (they would be observable or not depending on the proximity of a node). Another degree of freedom to probe the dissipation field and its temperature dependency would be to change the location of the heating point: it changes the shape of $T(x)$, thus can lead to further insight on the matter. This strategy has been used successfully to quantify the temperature field by tracking the resonance frequencies of several modes while scanning the heating position~\cite{Pottier-2021-JAP129}. The extension to thermal noise measurement still has to be demonstrated.

\medskip

The data that support the findings of this study are openly available in Zenodo~\cite{Fontana-2023-FDLink-Dataset}.

\medskip

\acknowledgments
We thank F. Vittoz and F. Ropars for technical support, A. Petrosyan, S. Ciliberto, B. Pottier and R. Pedurand for stimulating discussions, and V. Dolique, R. Pedurand, and G. Gagnoli of the Laboratoire des Matériaux avancés (LMA) in Lyon for the Tantala coating of sample C30C. Part of this research has been funded by the ANR project STATE (ANR-18-CE30-0013) of the Agence Nationale de la Recherche in France, and performed using the low noise chamber of the Optolyse platform, funded by the r\'egion Auvergne-Rh\^one-Alpes (CPER2016).

\appendix

\begin{figure*}[!t]
\begin{center}
\includegraphics[width=\textwidth]{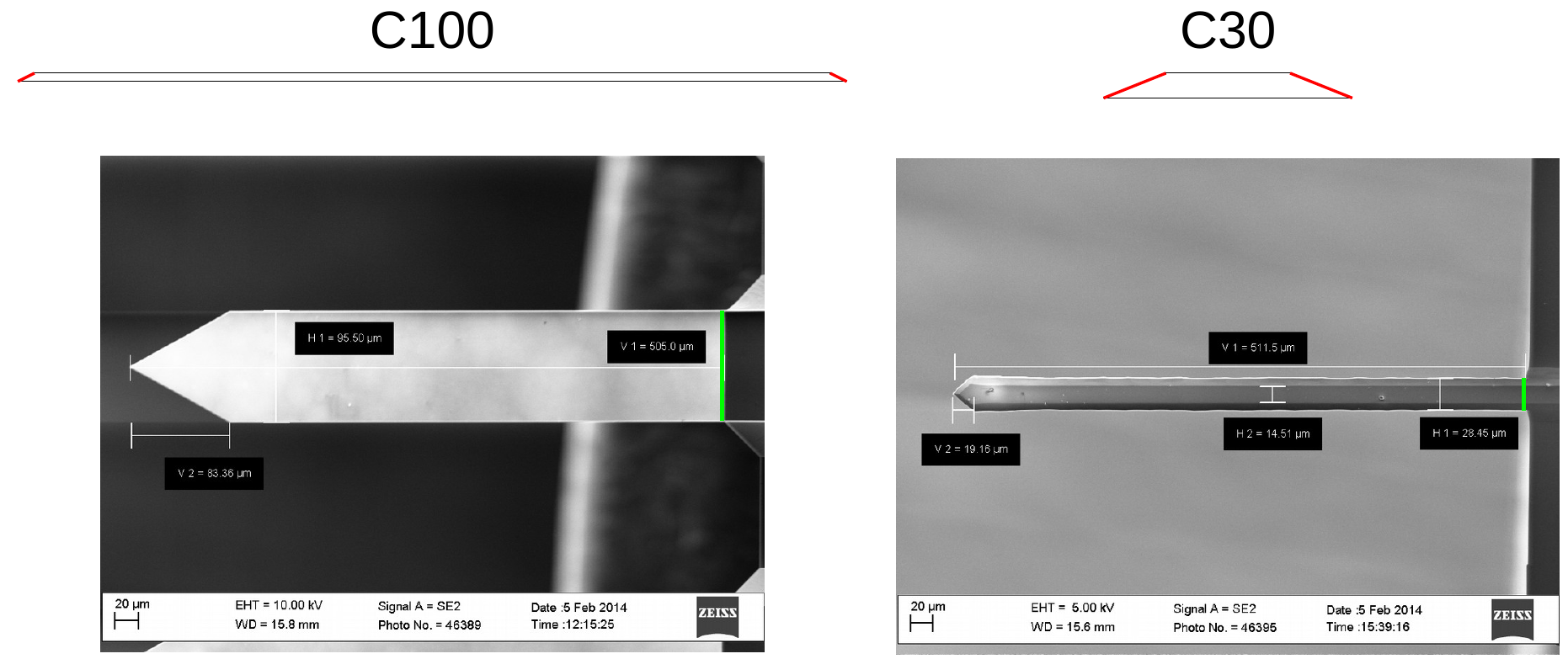}
\caption{Top: Cross section of samples C100 and C30, to scale. The red lines highlight the slanted faces of the cantilever. Bottom: Scanning electron microscopy image of the samples, with dimensions superposed. The green lines highlight the clamp length across the cantilever.}
\label{fig:crosssection}
\end{center}
\end{figure*}

\section{Geometrical differences between samples C30 and C100} \label{appendix}

It can be surprising at first to observe very different thermal noise behaviors on cantilevers C30 and C100, since both samples are made of the same material. However, after repeating the measurement on several cantilevers from different batches of both types, we came to the robust conclusion that they are demonstrating a different thermal noise amplitude and dissipation when heated. Though we have no strong insight on the manufacturing processes, doping of substrates used, thermal treatments, or thickness of the oxide layer, they actually have an important geometric difference. Indeed, the chemical etching which is usually used to manufacture the cantilever creates slanted sides, so that the cross section of the cantilever is trapezoidal instead of rectangular. This effect is barely noticeable on cantilever C100 ($\SI{1}{\mu m}$ thick, $\SI{100}{\mu m}$ wide), but much more on cantilever C30 ($\SI{2.7}{\mu m}$ thick, $\SI{30}{\mu m}$ wide). We plot in Fig.~\ref{fig:crosssection} the cross section to scale: cantilever C30 presents large areas which are slanted, and may contribute differently to the dissipation. Obviously also, the clamp between the cantilever and the base is approximately three times longer for cantilever C100, and enhances any clamp-induced dissipation.

\bibliography{FDlink}

\end{document}